# Thermo-osmosis in charged nanochannels: effects of surface charge and ionic strength

Wei Qiang Chen*, Andrey P Jivkov* and Majid Sedighi

School of Engineering, The University of Manchester, Manchester, M13 9PL, United Kingdom

*Corresponding authors: Weiqiang.Chen@manchester.ac.uk & Andrey.Jivkov@manchester.ac.uk

**Abstract**

Thermo-osmosis refers to fluid migration due to temperature gradient. The mechanistic understanding of thermo-osmosis in charged nano-porous media is still incomplete, while it is important for several environmental and energy applications, such as low-grade waste heat recovery, wastewater recovery, fuel cells, and nuclear waste storage. This paper presents results from a series of molecular dynamics simulations of thermo-osmosis in charged silica nanochannels that advance the understanding of the phenomenon. Simulations with pure water and water with dissolved NaCl are considered. First, the effect of surface charge on the sign and magnitude of the thermo-osmotic coefficient is quantified. This effect was found to be mainly linked to the structural modifications of aqueous electrical double layer (EDL) caused by the nanoconfinement and surface charges. In addition, the results illustrate that the surface charges reduce the self-diffusivity and thermo-osmosis of interfacial liquid. The thermo-osmosis was found to change direction when the surface charge density exceeds $-0.03 \text{C} \cdot \text{m}^{-2}$. It was found that the thermo-osmotic flow and self-diffusivity increases with the concentration of NaCl. The fluxes of solvent and solute are decoupled by considering the Ludwig–Soret effect of NaCl ions to identify the main mechanisms controlling the behavior. In addition to the advance in microscopic quantification and mechanistic understanding of thermo-osmosis, the work provides approaches to investigate a broader category of coupled heat and mass transfer problems in nanoscale space.

1. Introduction

Thermo-osmosis refers to the motion of a fluid due to a temperature gradient. The interest in this phenomenon in micro- and nano-scale spaces has recently increased due to its potential application in a wide range of fields. For example, in the area of low carbon energy conversion, nano-porous membranes are being explored for converting low-grade waste heat into mechanical energy, which in turn can be converted into electrical energy[1, 2] or in geo-materials where the transport of fluids in tight porous media under thermal effects is of importance in the energy geo-structures (e.g., energy foundations, geological disposal of radioactive waste).[3] In the emerging field of micro/nano-fluidics, thermo-osmosis is important to describe the fluid transport and mixing, with many promising applications such as the chip-level cooling without extra pumps,[4] and accurate control of (micro)nanoconfined cells, flexible fibers, colloids, etc. via shear flows.[5]

The understanding of thermo-osmosis in nano-porous media, including the underlying microscopic mechanisms, has been substantially improved as a result of recent theoretical, numerical, and experimental studies, e.g., ref. [2, 4-8]. Thermo-osmotic component of fluid flow and transport in many natural and technical systems occurs under the effects of surface charges and saline environment (e.g., charged nano-porous membranes for ionic selectivity,[9] osmotic energy recovery,[10] electric energy storage technologies,[11] and clay nanopores for saline geofluids are mostly negatively charged[12]). Thermo-osmosis in such systems is predominantly controlled by the pore size (nano-confinement effect), the charge of the solid surface (surface charge effect), and the salinity of pore fluid (ionic strength effect). Quantification of the nano-confinement effects on thermo-osmosis have progressed in literature.[6, 7] However, understanding the surface charge effects and ionic strength effects on thermo-osmosis is still limited, in contrast to the understanding of these two effects on other transport phenomena such as thermal diffusivity and Fickian diffusivity.[13-15] This paper aims to provide new insights into the surface charge effects and ionic strength effects on thermo-osmosis and the mechanisms controlling the phenomenon by isolating the coupled processes of thermo-osmosis and thermal diffusion (or Soret-Luwig effect).

The focus of this work is on: (1) thermo-osmosis of pure water confined by negatively charged silica nanochannels, where the pore fluid systems contain pure water and the surface charges are compensated by a small number of monovalent sodium counterions; and (2) thermo-osmosis of aqueous NaCl solutions confined by uncharged silica nanochannels, where the pore fluid systems contain NaCl with several concentrations. The choice of these systems was based on the fact that silica phases have become crucial constituents of many nanodevices for drug delivery,[16] water desalination,[17] and biomolecule detection.[18] The work addresses questions of long-standing scientific interest and current technological relevance, namely quantification of the interfacial structure, e.g., aqueous electrical double layer (EDL) on the charged surface, and the dynamical properties of interfacial liquid.[19, 20] Using non-equilibrium molecular dynamics (NEMD) simulations it is shown that the surface charges and ionic strength affect strongly the thermo-osmotic response of the interfacial liquid, which can be attributed to the alteration of aqueous electrical double layer (EDL).

2. **Molecular dynamics model**

Two MD systems have been studied in this work. The first system (system A) was designed to study the nanofluidic systems containing pure water and confined by charged silica substrates. **Fig. 1**a shows the first system that includes a nano-channel of silica with a channel size of 2.4 nm. The model contains around 1400 water molecules, 4400 silica surface atoms, and between 0 and 32 counterions (Na$^+$) depending on the surface charge density of the charged silica substrate. The system consists of pure water and a small number of counterions confined in a charged silica slit nanochannel. Periodic boundary conditions are imposed in all directions. It has been previously shown that the Onsager reciprocal relations are valid at the nanoscale,[2, 6, 7, 21] i.e., the thermo-osmotic coefficient relating the mass flux to the temperature gradient equals the mechano-caloric coefficient relating the heat flux to the pressure gradient. Because the application of pressure gradient is computationally simpler, and the steady-state flow can be achieved faster, the thermo-osmotic coefficient is determined here by calculating the mechano-caloric coefficient. A constant pressure gradient, $\nabla p = -\frac{1}{V}\sum_{i=1}^{N} f_i$, is created by applying an external particle force, $f_i$, in $x$ direction on every water molecule in the system with $N$ molecules (see **Fig. 1**a).

The second system (system B) was designed to study the effects of ionic solutes (NaCl) on thermo-osmosis. The system considered is neutral silica substrates containing aqueous NaCl solutions (see **Fig. 1**b). The concentrations studied are 0.00, 1.00, 1.23, 1.60, 2.29, 4.04 mol/kg.

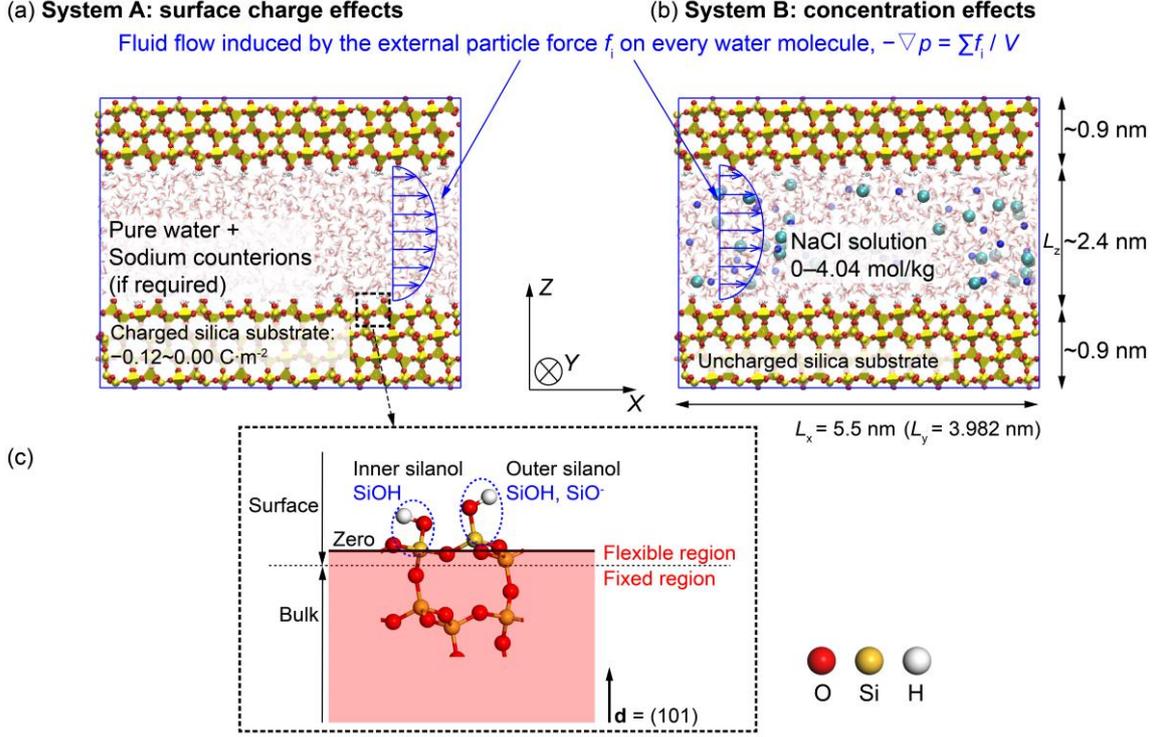

**Fig. 1.** Illustration of the MD systems under pressure gradient. System A contains pure water confined by charged silica surfaces (a). System B contains NaCl aqueous solution confined by uncharged silica surfaces (b). The silica surface with the outer and inner silanol groups, where the zero plane is shown by the solid black line, and the dashed black line distinguishes surface silica atoms from bulk ones (c).

Upon applying the constant pressure gradient, the system reaches steady state with a constant planar Poiseuille flow after a period of equilibration time. The velocity profiles $v_x(z)$ at steady-state are recorded and fitted to the Stokes equation[22]

$$v_x(z) = \frac{-\nabla p}{2\eta}\left[db + \left(z + \frac{d}{2}\right)\left(\frac{d}{2} - z\right)\right] \quad (1)$$

where, $d = L_z - 2z_s$ is the actual channel width after deducting the thickness of immobile layers $z_s$. $\eta$ and $b$ are the fitted liquid viscosity and interlayer slippage length, respectively.

The thermo-osmotic coefficient $m_{21}$ is computed by[6, 7]

$$m_{21} = -\frac{J_h}{\nabla p} \tag{2}$$

where, $J_h$ is the heat flux. It is calculated from the excess specific enthalpy $\delta h(z)$ and the velocity profile $v_x(z)$ by

$$J_h = \frac{1}{L_z}\int \delta h(z)v_x(z)dz \tag{3}$$

where, the specific excess enthalpy is calculated by

$$\delta h(z) = h(z) - h_b \tag{4}$$

where, $h_b$ is the specific enthalpy of the bulk liquid, and $h(z)$ is the local specific enthalpy. The latter is calculated by[23]

$$h(z) = p_{xx}(z) + u(z) \tag{5}$$

where, $p_{xx}(z)$ is the calculated pressure component of interfacial liquid along the x-direction using the atom-based virial equation,[24] and $u(z)$ is the specific internal energy (per unit volume) of interfacial liquid. For sufficiently wide channels $h_b = h(z = 0)$.[2, 23, 25]

Considering the hydrodynamic interlayer slippage length, $b$, and the thickness of the immobile layer, $z_s$, the theoretical value of thermo-osmotic coefficient, $M_{21}$, can be obtained by[2]

$$M_{21} = \frac{1}{\eta}\int_{z_s}^{+\infty}(z - z_s + b)\delta h(z)dz \tag{6}$$

The difference between $m_{21}$ and $M_{21}$ is that the former uses the measured velocity profile of interfacial liquid, while the latter assumes a linear velocity profile in the boundary layer with the same hydrodynamic boundary condition as the former.

The EDL is the origin of many electrokinetic transport phenomena at charged interfaces such as electro-osmosis, electrophoresis, streaming potential, and streaming current. Like the streaming current experiments, the induced fluid flow of interfacial liquid allows for quantifying the widely used zeta potential, $\zeta$, which is defined as the electrostatic potential at the position of the plane of shear. The zeta potential is an important quantity in electrokinetic transport because it quantifies the coupling between electrostatic properties of the ion cloud and hydrodynamics of the solvent within the EDLs. Following previous studies[26, 27] and using the Helmholtz–Smoluchowski equation, it can be computed by

$$\zeta = -\frac{J_e}{(-\nabla p)}\frac{\eta_b}{\varepsilon_b} \qquad (7)$$

where, $J_e = \frac{1}{L_z}\int \rho_q(z)v_x(z)dz$ is the streaming electric current density, $\rho_q(z)$ the charge density, $-\nabla p$ the applied pressure gradient, $\eta_b$ and $\varepsilon_b$ are the bulk viscosity and dielectric permittivity of the liquid, respectively.

Similarly, the theoretical prediction of the zeta potential, $\zeta^*$, can be obtained by[28, 29]

$$\zeta^* = \frac{1}{\varepsilon_b}\int_{z_s}^{+\infty}(z - z_s + b)\rho_q(z)\mathrm{d}z \qquad (8)$$

## 3. Molecular dynamics simulations

All NEMD simulations were implemented with the LAMMPS.[30] The Velocity-Verlet integration is used to solve the Newton's equations of motion with a time step of 3.0 fs.

**Silica Substrates.** The silica substrates with surface charge densities of $0.00$, $-0.03$, $-0.06$, and $-0.12\ \mathrm{C\cdot m^{-2}}$ were built following our previous work.[15] The partial charges on atoms were determined based on the study of Kroutil et al.[31] The interatomic interactions were computed by the ClayFF force field.[32]

**Nanoconfined liquid.** Sufficient water molecules were packed into the silica nanochannels and 0–32 sodium counterions with positive charges were dissolved in the interfacial water to compensate the negatively charged silica surfaces. For system B, NaCl ions were further dissolved into the interfacial liquid to achieve the specified ion concentrations. The formed interfacial liquids were described with the rigid SPC/E water model[33] and the force field by Dang et al.[34] for the ion-ion(water) interactions. The Coulombic and Lennard-Jones 12-6 potentials are adopted to describe the silica-liquid interactions and standard Lorentz-Berthelot mixing rules were used to determine the cross-interaction parameters between various atomic types. A cut-off radius of $r_c = 1.5 \text{nm}$ is used for short-range interactions, and the long-range electrostatic interactions were determined by the Particle-Particle-Particle-Mesh (PPPM) method.[35] The bonds involving hydrogen atoms in the system were constrained during simulation to enable a longer timestep, i.e., 3.0 fs. The same large time step of 3 fs, together with increased interatomic force cut-off, have been used by some previous MD studies, e.g., ref. [36], to perform NEMD simulations of rigid SPC/E water. In our work, the bulk atoms of silica slabs are tethered to their respective initial positions by independent springs during simulation, and the surface atoms are flexible with constrained O-H bonds in silanol groups to enable a longer timestep. We repeated the simulations for several cases with a smaller time step of 1 fs. The results showed a negligible difference, therefore a time step of 3 fs was used.

**NEMD Simulations.** An energy minimization was first performed on the established system using the steepest descent algorithm. A relaxation stage composed of a 0.1 ns run in *NVT* ensemble and 1 ns run in *NPT* ensemble, using Nose-Hoover thermostat and barostat, was then carried out to achieve a system pressure of $P = 600 \text{ bar}$ with a coupling time of 3 ps, and a system temperature of $T = 300 \text{ K}$ with a coupling time of 0.3 ps. The adopted thermodynamic conditions are consistent with our previous study[15] on the thermal diffusion of aqueous NaCl solutions occurring at the same nanofluidic system. This enabled to directly use the thermal diffusion coefficient obtained in that study and further decouple the fluxes of solvent and solute, see details in the Discussion Section. Following the previous studies and simulation procedures of Kroutil et al. [31] and Quezada et al. [37] on the same systems where silica surfaces are immersed in aqueous NaCl solutions, the semi-isotropic pressure coupling condition is used in the *NPT* simulation run, where the Nose-Hoover barostat set to 600 bar was used with scaling only in the z-direction of the simulation box, which is due to the strong structural heterogeneity of the simulation box in the $z$ direction and $x/y$ direction.

The simulation was followed by applying the external particle force and creating a pressure gradient equalling to $-\nabla p = 0.055 \text{GPa/nm}$ inside the nanochannel, and an equilibrium run was performed. The steady state was achieved after approximately 9 ns. Finally, a production run was performed for 18 ns. The system configuration was recorded and analysed every 0.3 ps to produce thermodynamic data, density, charge, excess enthalpy, velocity profiles, etc., by dividing the simulation cell into typically 600 sampling layers along the $z$ direction. During the equilibrium and production runs, the system temperature was modulated by a canonical sampling thermostat adopting global velocity rescaling with Hamiltonian dynamics.[38] The surface atoms (see **Fig. 1**c) were thoroughly flexible throughout the simulation. In the *NVT* simulations, bulk atoms were constrained in all directions (see **Fig. 1**c), while in the *NPT* simulations, they were permitted to move only in the $z$ direction to enable the pressure equilibration. These fixations were achieved by applying a spring with force constant of $1000 \text{ kJ mol}^{-1} \text{ nm}^{-2}$ to tethered atoms in the specified directions. For better comparison between different cases, the zero line was specified as the averaged position of all surface silicon atoms (see **Fig. 1**c).

## 4. Results

### 4.1 Structural properties of nano-confined liquid

**Surface charge effect.** For every simulation case of the system A, the mass/number and charge density profiles of water and $\text{Na}^+$ as a function of distance from the silica surface were calculated. The results are shown in **Fig. 2**. The zero plane of silica surface defined in **Fig. 1**c is fixed at the abscissa $x = 0$ for clearer comparison. The molecular organization of the interfacial water into layers near the silica surface can be clearly identified by the peaks and valleys shown in **Fig. 2**a–b. This layering becomes increasingly pronounced with decreasing surface charge density. Cation number density near the silica surface in **Fig. 2**a shows that $\text{Na}^+$ form three categories of adsorbate species: (i) inner-sphere surface complexes (ISSC), (ii) outer-sphere surface complexes (OSSC), and (iii) diffuse swarm (DS) species.[39] The corresponding distinct counterion adsorption planes, namely, 0-plane, β-plane, and d-plane of the widely used triple layer model (TLM[40]) can be clearly identified. These are depicted by blue vertical lines in **Fig. 2**a–c. Further, **Fig. 2**a–b shows that with decreasing surface charge density more water molecules and sodium counterions are attracted to the charged surfaces, forming ISSC and OSSC with different ionic solvation structures and less mobility compared with those in the bulk liquid, where ISSC is partially hydrated and adsorbed directly on the silica surfaces and OSSC is fully hydrated. The accumulation of sodium counterions on the surface increases with the surface charge and is limited mainly to ISSC and OSSC regions, where the local viscosity will increase accordingly. An enhanced local fluid viscosity in the EDLs near the solid surface compared with that in the bulk fluid away from the solid surface has been reported by previous studies, e.g., ref. [41, 42]. This results in an apparently enhanced effective viscosity of the interfacial fluid. It was also previously found that increasing the surface charge density can further increase this apparent viscosity, due to the increasing adsorption and entrapment of counterions on charged surfaces, and resultant lateral hindrance for the fluid flow of surrounding solvent molecules.[41] Similar observation will be made in section 4.3.

**Fig. 2**b shows the number density of water hydrogen atoms ($H_w$) and water oxygen atoms ($O_w$), which indicates that the water mass density in **Fig. 2**a is dictated by the number density of water oxygen atoms due to its much larger mass relative to the water hydrogen atoms. **Fig. 2**c shows the charge density distribution of interfacial liquid, which can be used to evaluate the electro-kinetic responses at the interface in the subsequent analysis. The results indicate that the charge density distribution is concurrently determined by the distribution of sodium counterions, water hydrogen atoms, and water oxygen atoms, and its peaks and valleys are consistent with those in **Fig. 2**b. The charge density is zero in the bulk liquid due to the random arrangement and distributions of water molecules and sodium counterions.

The thickness of the interfacial EDLs (including Stern layer and diffuse ion swarm in **Fig. 2**a) can be characterized by the Debye length, $\lambda_D$. Within the classical mean-field theories and for dilute electrolyte (low ion density) regime, the Debye length can be calculated by[43]

$$\lambda_\mathrm{D} = \left(\frac{\varepsilon_\mathrm{r}\varepsilon_0 k_\mathrm{B} T}{\sum_i \rho_{\infty i} e^2 z_i^2}\right)^{\frac{1}{2}} = \left(\frac{1}{4\pi l_B \sum_i \rho_{\infty i} z_i^2}\right)^{\frac{1}{2}} \tag{9}$$

where, $\varepsilon_r$ is relative dielectric permittivity of interfacial fluid, $\varepsilon_0$ the vacuum permittivity, $k_B$ the Boltzmann's constant, $T$ the temperature, $\rho_{\infty i}$ the number density of ion type $i$ in the bulk region in **Fig. 2**a, $e$ the elementary charge, $z_i$ the ion valency, and $l_B = \frac{e^2}{4\pi\varepsilon_r\varepsilon_0 k_B T}$ is the Bjerrum length. Substituting $\varepsilon_r = 70.8^{44}$ into Eq. (9) gives $\lambda_\mathrm{D}$ in the range of 0.49–0.67 nm for charged silica surfaces in **Fig. 2**a. This theoretical screening length is in good agreement with the one obtained by MD simulations.

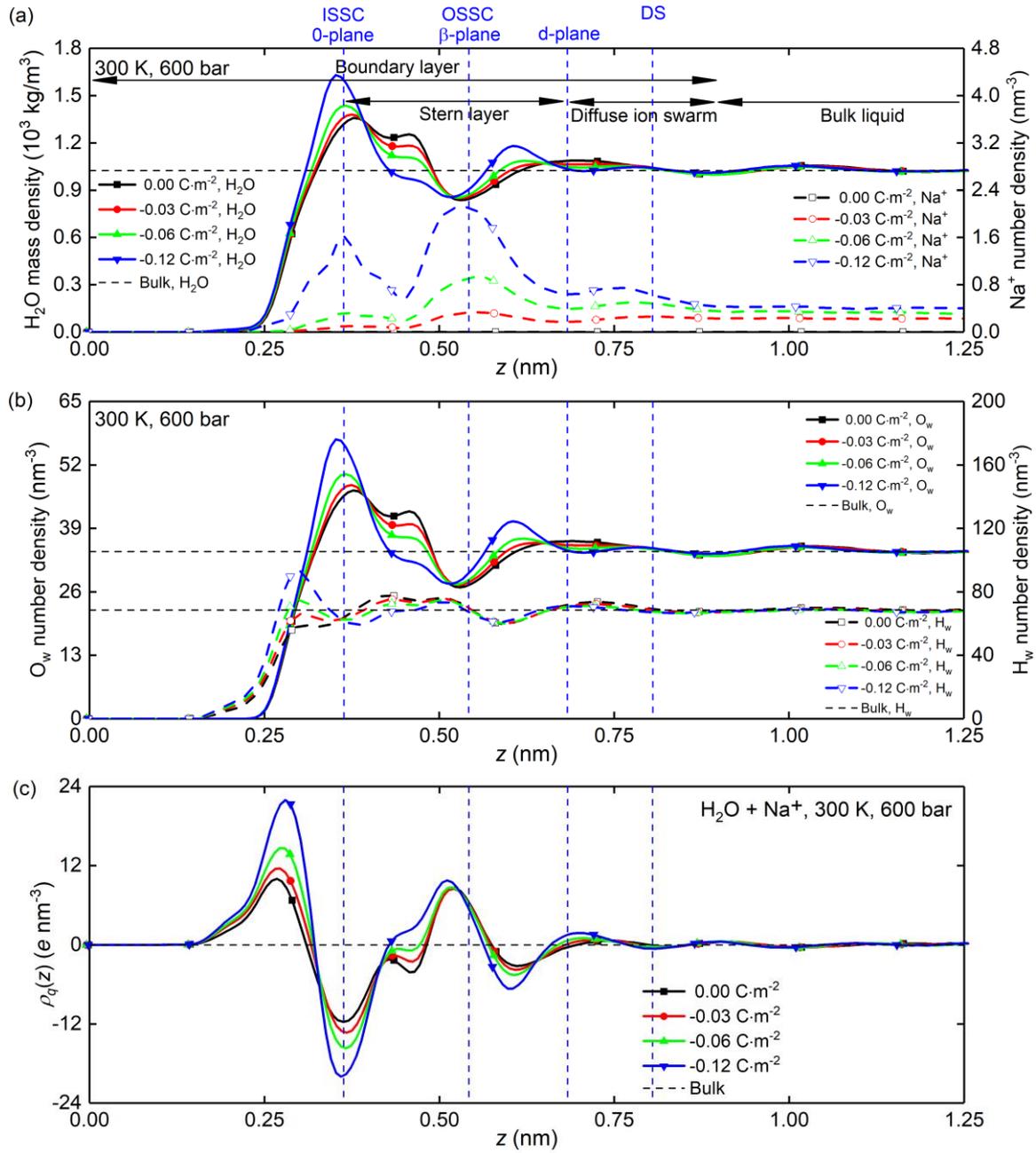

**Fig. 2.** The structural properties of system A: (a) Mass density of water molecules and number density of sodium counterions; (b) Number density of water oxygen atoms and water hydrogen atoms; and (c) Charge density of interfacial liquid. All data corresponds to the silica nanochannel of width $L_z = 2.4$nm.

**Concentration effect.** The EDL structure of the interfacial liquid for system B is presented in **Fig.**

**3**. **Fig. 3**a–c show that the shape of number/mass density profile of water molecules and NaCl ions, and the position of TLM planes are hardly modified by the addition of NaCl ions. This is confirmed by the unchanged positions of peaks and valleys in the profiles and the vertical blue dashed lines. Accordingly, the charge density distribution in **Fig. 3**d is also kept constant with the addition of NaCl ions. In addition, **Fig. 3**c shows an obvious cation condensation and anion exclusion (repulsion) at the silica surface, where the number density distribution profile of chloride ions roughly starts beyond the 0-plane. The distribution of sodium and chloride ions becomes equal in the bulk liquid. The increasing concentration also leads to enhanced ion pairing between sodium and chloride ions, which might concurrently decrease the mobility of each ion.

Note that in the system B, where concentrated aqueous NaCl solutions (beyond the Debye−Hückel region) are confined by two planar charged surfaces, a previous experimental study[45] with similar setup has pointed out that in such concentrated electrolytes, the theoretical Debye screening length of Eq. (9) will become inapplicable and substantially underestimate the real thickness of EDL, i.e., the electrostatic screening length, which was also found to increase with concentration. This is also consistent with our MD results in **Fig. 3**, where $\lambda_D$ is estimated to be only within 0.13–0.26 nm according to Eq. (9).

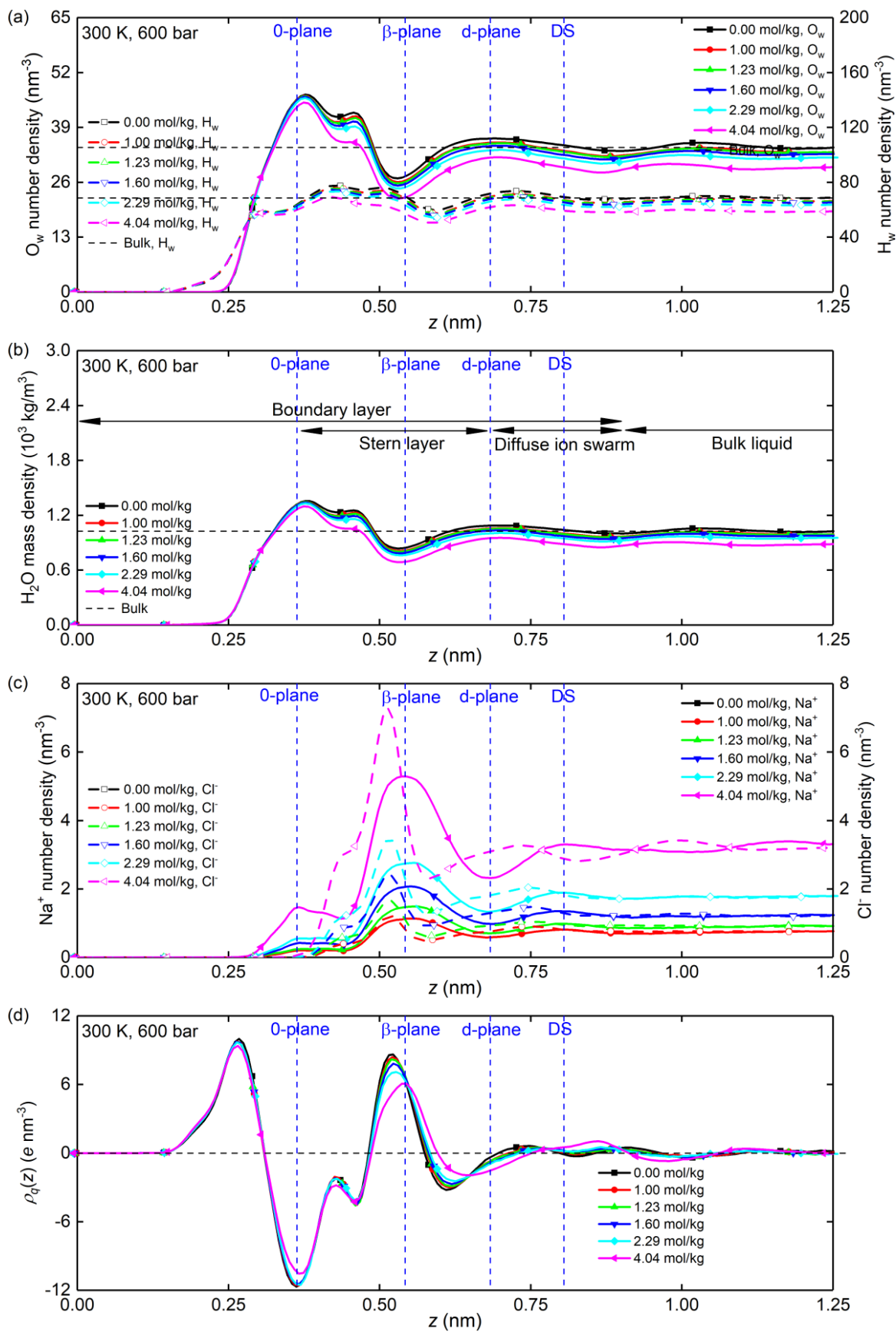

**Fig. 3.** The structural properties of system B: (a) The number density distribution of water hydrogen atoms and water oxygen atoms; (b) The mass density distribution of water molecules; (c) The number density distribution of sodium ions and chloride ions; and (d) The charge density distribution of interfacial liquid. All data corresponds to the silica nanochannel of width $L_z = 2.4$nm.

### 4.2 Thermodynamic properties of nanoconfined liquid

**Surface charge effect.** For system A, the pressure component in the $x$ direction and the excess specific enthalpy of interfacial liquid are shown in **Fig. 4**a and b, respectively, where the latter is calculated by Eq. (4). The regions of the boundary layer and the bulk liquid are consistent with those in **Fig. 2**a. The excess enthalpy profiles in **Fig. 4**b oscillate notably in the boundary layers and diminish to zero in the bulk liquid. It is found that the excess specific enthalpy tends to more negative with decreasing surface charge density, due to the enhanced attractive interactions relative to the weakened repulsive interactions between interfacial liquid particles and the negatively charged surfaces,[46] e.g., the electrostatic attractions between sodium counterions and surfaces. Following previous studies,[6, 23, 47] thermo-osmotic forces acting on interfacial liquid under a unit temperature gradient, i.e., $-\nabla T = 1$, are calculated and shown in **Fig. 4**c and d, where the former is the pressure on the fluid element, $f_x(z) = -\delta h(z)\frac{\nabla T}{T}$, and the latter is the body force on liquid particles, $f_x^P(z) = \frac{f_x(z)}{\rho_N(z)}$, where $\rho_N(z)$ is the number density distribution of interfacial liquid along the $z$ direction. It is found that the thermo-osmotic force driving the fluid flow emerges only in the non-bulk regions defined in **Fig. 2**a, in either $+x$ or $-x$ direction depending on its $z$ position. According to the TLM defined in the previous section, it is found that the interfacial liquid located in β-plane and d-plane is thermophilic, i.e., the liquid will migrate from colder to hotter areas under a temperature gradient, while the liquid in 0-plane is thermophobic, i.e., the liquid will migrate from hotter to colder areas. In addition, **Fig. 4**c and d show that the thermo-osmotic force acting on interfacial liquid tends to be more negative with decreasing surface charge density, which indicates that thermo-osmotic flow will tend to $-x$ direction when the surface charge density is applied. This tendency can be presented more clearly by comparing the integrals

$\int_{-L_z/2}^{L_z/2} f_x(z)\mathrm{d}z$ and $\int_{-L_z/2}^{L_z/2} f_x^P(z)\mathrm{d}z$ in **Fig. 4**c and d, which decrease with increasing surface charge density. In fact, the effective excess enthalpy, $\int_{-L_z/2}^{L_z/2} \delta h(z)\mathrm{d}z$, has been used in many previous studies, e.g., [5], to determine the direction of thermo-osmosis, as is shown in **Fig. 4**b. This is the same as our integrals of thermo-osmotic forces. Therefore, it can be expected that the thermo-osmotic coefficient may become negative with decreasing surface charge density. This expectation will be proved shortly.

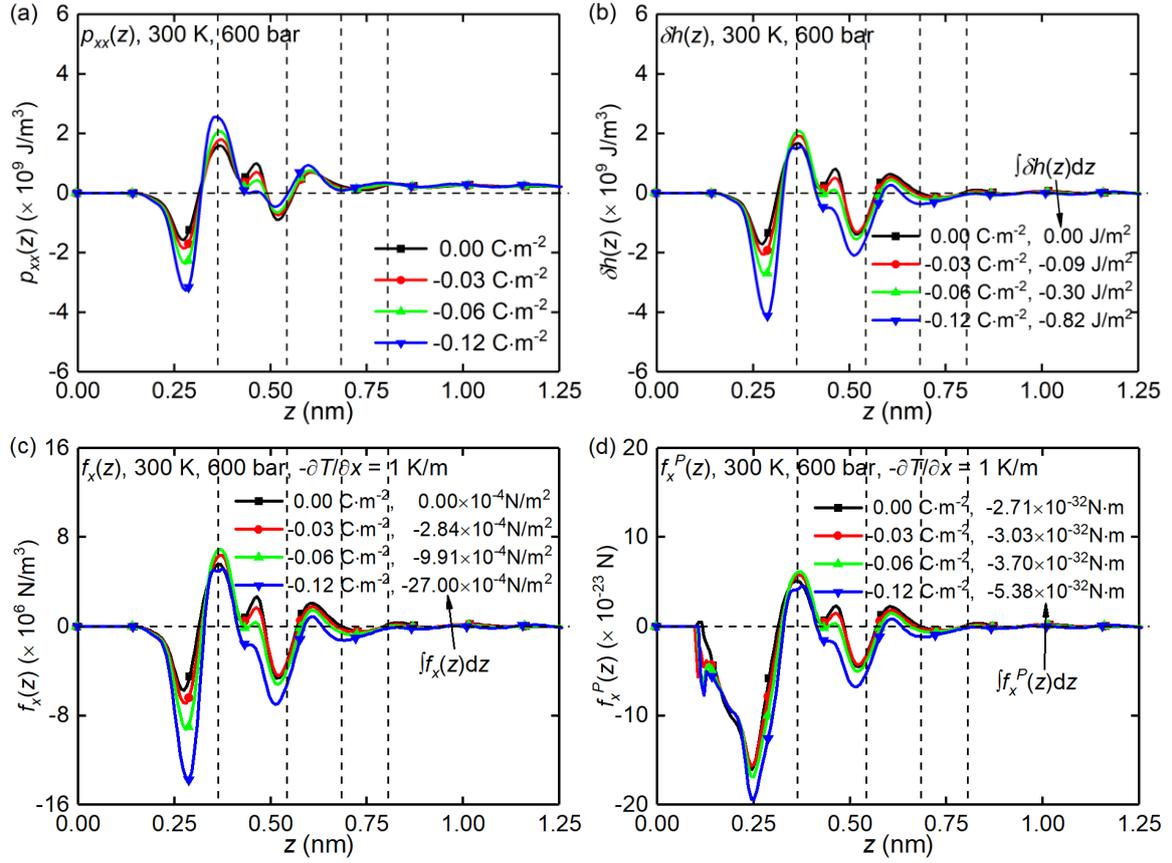

**Fig. 4.** System A distributions of: (a) pressure component along the $x$ direction; (b) excess specific enthalpy; (c) body force on liquid element; and (d) body force on liquid particle induced by a unit temperature gradient. All data corresponds to the silica nanochannel of width $L_z = 2.4$ nm.

**Concentration effect.** For system B, the thermodynamic properties of interfacial liquid are computed and presented in **Fig. 5**. It is found from **Fig. 5**a–b that $p_{xx}(z)$ generally keeps constant with the addition of NaCl ions, while $\delta h(z)$ varies significantly, indicating that the internal energy of interfacial liquid changes a lot with the addition of NaCl ions according to Eq. (5). **Fig. 5**c–d shows that the addition of NaCl ions increases the magnitude of the thermo-osmotic force on the interfacial liquid either in the $+x$ or $-x$ direction. The position of these enhancements is consistent with that of TLM planes, indicating that the addition of NaCl ions facilitates the thermo-osmosis either in the $+x$ or $-x$ direction depending on their occupied TLM planes. Further, the integral of the thermo-osmotic force over the channel section monotonically increases with the addition of NaCl ions, indicating an overall flowing trend towards the $+x$ direction.

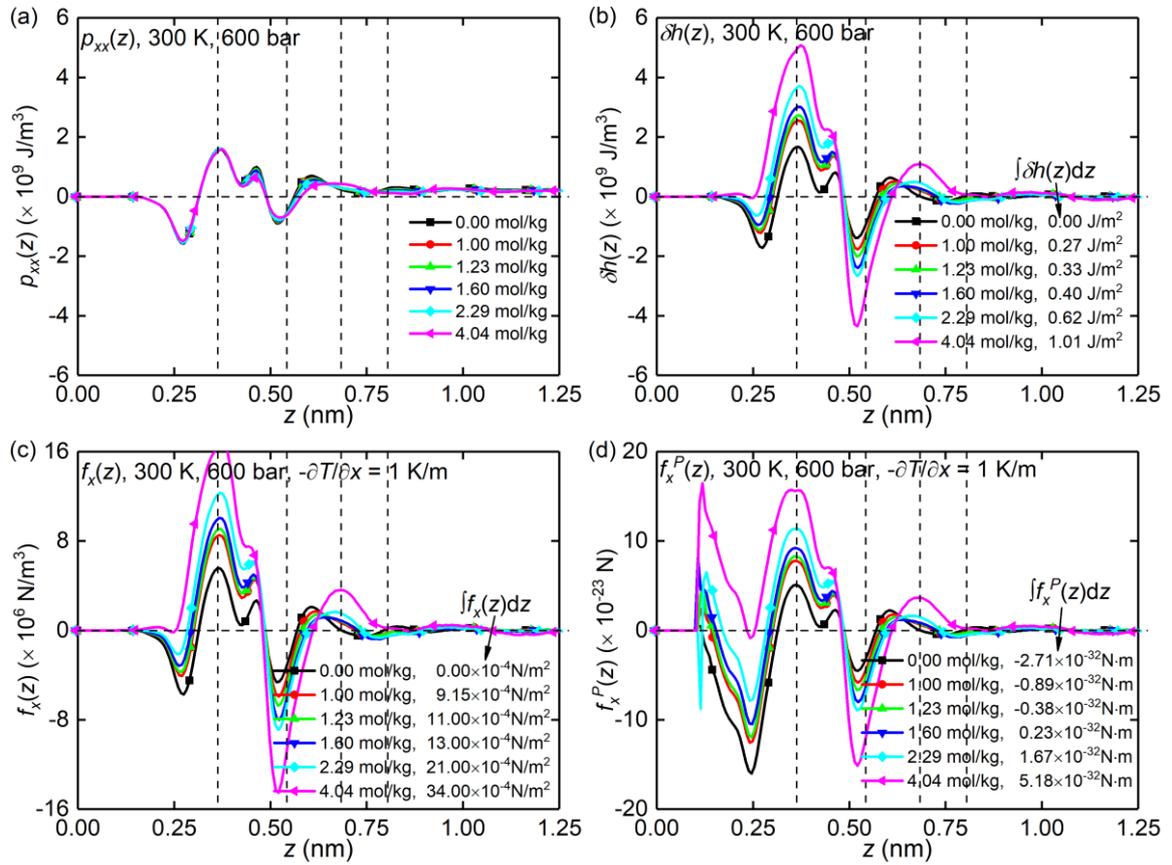

**Fig. 5.** System B distributions of: (a) pressure component along the $x$ direction; (b) excess specific enthalpy; (c) body force on liquid element; and (d) body force on liquid particle induced by a unit temperature gradient. All data corresponds to the silica nanochannel of width $L_z = 2.4$ nm.

**4.3 Hydrodynamic properties of nanoconfined liquid**

**Surface charge effect.** For system A and by applying an external particle force on each water molecules, the generated Poiseuille-type parabolic velocity profiles of interfacial liquid confined by silica surfaces are plotted in **Fig. 6**b–e. These can be described well by the continuum hydrodynamics Eq. (1), even in the range of EDLs. **Fig. 6**a shows the definitions of the fitting parameters in Eq. (1), where the shear plane position, $z_s$, in which the slip boundary condition is imposed, is determined as the positions of first peaks of the water density profiles in **Fig. 2**a. The derived fitting parameters in **Fig. 6**f show that with decreasing surface charge density, the viscosity of confined liquid increases and the thickness of (stagnant) immobile layers decreases. These observations are consistent with the results in the previous subsection that more counterion complexes are formed near the charged silica surfaces with decreasing surface charges. The vertical dashed lines in **Fig. 6**b–e indicating the TLM planes show that the interfacial liquid with less mobility, i.e., being nearly stagnant, is located between the silica surfaces (zero line in **Fig. 1**c) and 0-planes. The description of immobile layer is from classical continuum theories, which define a fluid adsorption layer near the solid surface with infinite fluid viscosity and zero fluid mobility.[42] Notably, the MD data in **Fig. 6** suggests that the fluid molecules located within the defined immobile layer is nearly immobile, but not completely immobile. Due to the strong solid-fluid friction and higher local fluid viscosity, they move relatively slowly and nearly synchronically as is observed by the relatively flattening of velocity profiles. These findings are consistent with a previous MD study of Moh et al.[48] on the decane transport through slit calcite pores. In addition, due to the hydrophilic nature of the silica surface,[6,7] the interlayer slippage length, i.e., the velocity jump, is small, and its magnitude decreases with the surface charge density becoming more negative. This is consistent with a previous study on the interface between aqueous sodium chloride and charged graphene.[26] The decreasing interlayer slippage length can be attributed to the enhanced binding (trapping) effect of $Na^+$ counterions on the negatively charged sites of the silica surfaces with decreasing surface charge. The trapped counterions stick out from the silica surface inducing a viscous Stokes drag.

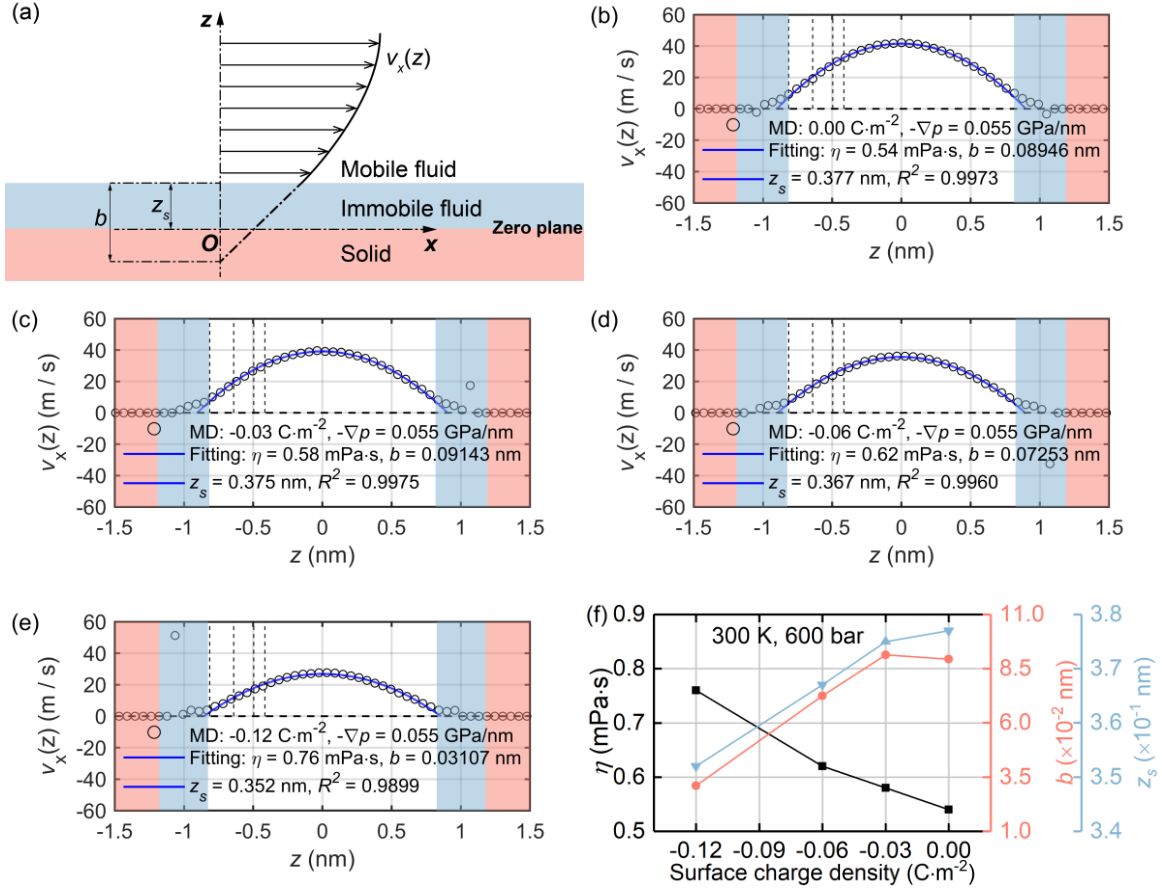

**Fig. 6.** Hydrodynamic properties of System A: (a) Illustration of liquid velocity profile normal to a solid surface, where the shear plane position $z_s$ is shown; (b–e) Velocity profiles of interfacial liquid along $x$-direction under different surface charge densities, where the red and blue regions correspond to those defined in the subfigure a, and vertical dashed lines indicate different planes defined in **Fig. 2**; and (f) The variations of viscosity, thickness of immobile layer, and interlayer slippage length of interfacial liquid with the surface charge density.

**Concentration effect.** For system B, by applying an external particle force on every water molecule, the generated velocity profiles of interfacial liquid for different cases are presented in **Fig. 7**a–f, which can be well described by the continuum hydrodynamics. The fitting parameters in **Fig. 7**g show that increasing NaCl concentration leads to monotonically increasing liquid viscosity $\eta$ and interlayer slippage length $b$, with the thickness of immobile layer $z_s$ being almost constant. Overall, a general decrease of liquid mobility can be observed when the concentration increases. Note that even though the channel size is fixed at around 2.4 nm to exclude the channel size effects, whereas the addition of NaCl ions may slightly enlarge the channel size, e.g., see **Fig. 7**a–f.

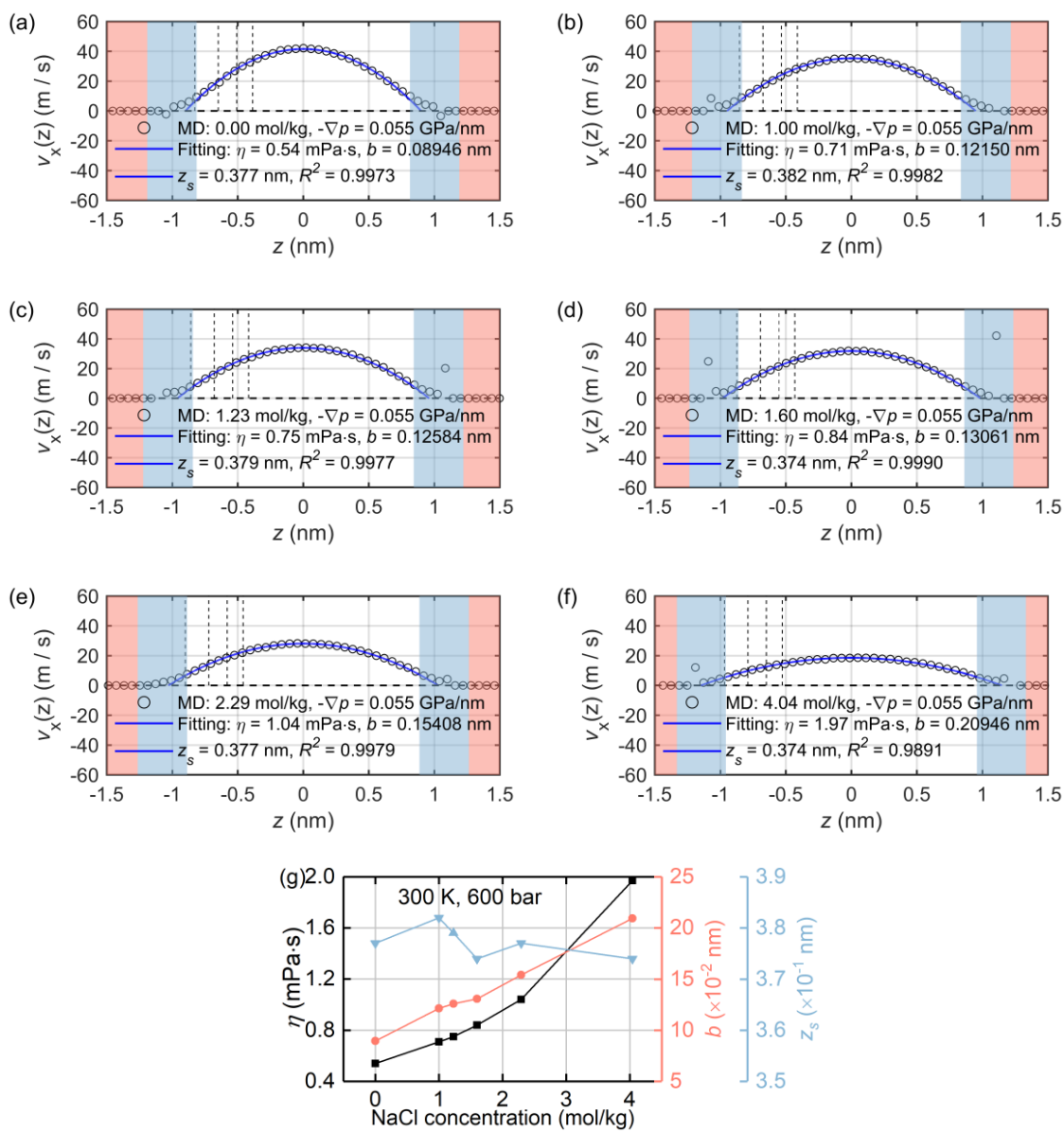

**Fig. 7.** Hydrodynamic properties of System B: (a–f) The velocity profiles of interfacial liquid along $x$-direction under different NaCl concentrations, where the vertical dashed lines indicate different TLM planes; and (g) The variations of viscosity, thickness of immobile layer, and interlayer slippage length of interfacial liquid with the NaCl concentration.

**Self-diffusivity of interfacial liquid.** The self-diffusion coefficient is an important quantity, which can be used to estimate the transport distance of mass diffusion within a given time, e.g., ref. [49]. For system A and B, the one-dimensional self-diffusion coefficient of interfacial liquid, including water molecules (at the center of mass) and NaCl ions (if exist), in the $x$ direction ($D_x$) were calculated for the entire interlayer space by using the Einstein relation:[50]

$$D_x = \lim_{\tau \to \infty} \frac{1}{2d} \frac{d[(MSD_x(\tau)]}{d\tau} \tag{10}$$

where, $d = 1$ is the diffusion dimensionality, $\tau$ is the simulation time, and $MSD_x(\tau) = \langle (x(\tau) - x(0))^2 \rangle$ is the mean-square displacement of interfacial liquid along the $x$ direction.

The self-diffusivity of interfacial liquid under bulk conditions is determined for comparison by averaging the mean-square displacements over three dimensions and setting $d = 3$. The dependance of $D_x$ on the surface charge density and ionic strength are presented in **Fig. 8**a and b, respectively. It can be observed from **Fig. 8**a, that both the nanoconfinement and the surface charges can reduce the self-diffusivity of interfacial liquid. The obtained bulk value, i.e., $(2.91 \pm 0.01) \times 10^{-9} m^2/s$, is consistent with the previous MD study for the same rigid SPC/E water model under similar thermodynamic condition, i.e., $(2.97 \pm 0.05) \times 10^{-9} m^2/s$.[51] This provides further confidence in the accuracy of the simulations presented here. Furthermore, **Fig. 8**b shows that the increasing ionic strength can also reduce the self-diffusivity of interfacial liquid. The same varying trends of interlayer diffusion with increasing surface charge density and ion concentration have also been reported by Greathouse et al. [49] for MD systems consisting of montmorillonites and interlayer aqueous NaCl/CaCl$_2$ solutions. They explained that a higher surface charge density makes interlayer environment more hydrophilic with slower diffusivity of interfacial liquid, which well agrees with our results in **Fig. 8**a, and **Fig. 6**f. Since the diffusion coefficient is inversely proportional to the viscosity according to Stokes–Einstein relation, our results for interlayer self-diffusivity in **Fig. 8** agree well with the calculated viscosity in **Fig. 6**f and **Fig. 7**g.

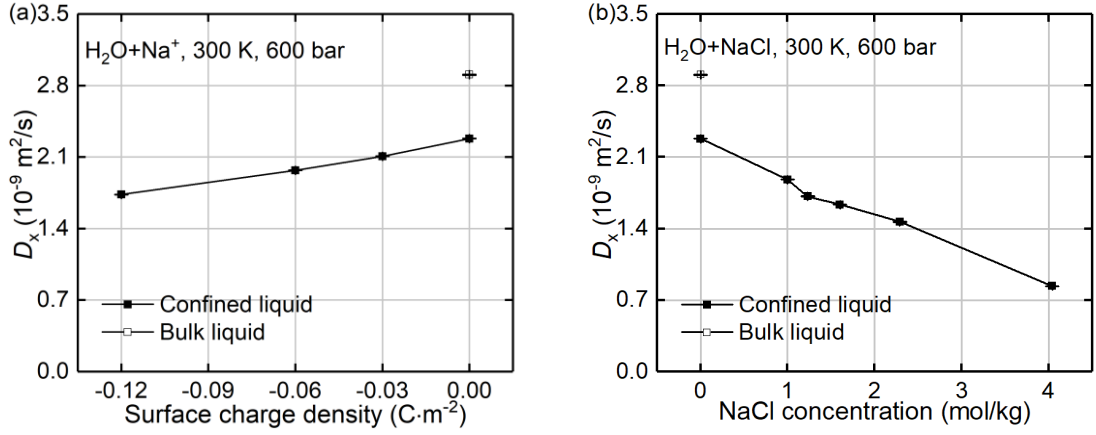

**Fig. 8.** The self-diffusion coefficient of interfacial liquid under different (a) surface charge densities, and (b) NaCl concentrations.

### 4.4 Thermo-osmotic properties of nanoconfined liquid

**Surface charge effect.** For system A, the thermo-osmotic coefficient $m_{21}$ is calculated by Eq. (2). **Fig. 9**a shows the variation of thermo-osmotic coefficient with surface charge density. Three independent simulations with various initial states of interfacial liquid were performed to determine the variability of the computed $m_{21}$. The variabilities are shown by the error bars in the figure. It was found that decreasing surface charge can significantly reduce the thermo-osmotic coefficient. The coefficient changes sign to negative when the surface charge density is below $-0.03\, C\cdot m^{-2}$, indicating that the thermo-osmotic flow would change its direction and liquid would migrate towards the hot region. At this threshold, there is no thermo-osmotic flow. This conclusion is consistent with the analysis of thermo-osmotic force in **Fig. 4**. Note that the calculated thermo-osmotic coefficients are comparable to those from previous experiments[52, 53] on silica-based materials, $\pm(10^{-10} \sim 10^{-9})\, m^2/s$, and that both positive and negative signs of the thermo-osmotic coefficient have been observed in those experiments. Bregulla et al.[52] have attributed the negative thermo-osmotic coefficient to the electrostatic contribution of electrical double layer caused by surface charges. In addition, it was found that the thermo-osmotic coefficient $(10^{-10} \sim 10^{-8}\, m^2/s)$ is of the same order of magnitude or larger than the self-diffusion coefficient $(10^{-10} \sim 10^{-9}\, m^2/s)$. This shows that the thermo-osmotic flow at nanoscale ought not be overlook when considering coupled transport. Substituting the corresponding values in **Fig. 4**b and **Fig. 6**f into Eq. (6), the computed $M_{21}$ is presented in **Fig. 9**a. This is found to agree well with the calculated $m_{21}$. The discrepancies come from the linear assumption for the velocity profile in the boundary layers when deriving the Eq. (6).[2] A high consistency between $m_{21}$ and $M_{21}$ is found at low surface charge densities, i.e., below $-0.03 \text{C} \cdot \text{m}^{-2}$, while their discrepancy increases with the increasing surface charge density.

**Concentration effect.** The thermo-osmotic response for different cases in system B is quantified. The thermo-osmotic coefficient, the measured $m_{21}$ and the predicted $M_{21}$, are presented in **Fig. 9**b. It is found that the measured thermo-osmotic coefficient $m_{21}$ generally increases with the addition of NaCl ions, though it experiences a slight drop at the concentration of around 1.6 mol/kg. The predicted one, $M_{21}$, has a trend consistent with $m_{21}$ but with a lower magnitude. It is found that the increasing NaCl concentration will enlarge the difference between $m_{21}$ and $M_{21}$. This is consistent with the effect of surface charge density on the difference between $m_{21}$ and $M_{21}$ as is presented in **Fig. 9**a. Similarly, the difference is due to the linear assumption for the velocity profile in the boundary layers when $M_{21}$ is derived. According to the formulas for $m_{21}$ and $M_{21}$, the thermo-osmotic response is the combined action of thermodynamic properties in **Fig. 5**b and hydrodynamic properties in **Fig. 7**g, where the magnitude of the former one increases with increasing NaCl concentration, either in $+x$ or $-x$ direction, while the latter one decreases with increasing the NaCl concentration. Their competition finally determines the variation of the thermo-osmotic coefficient seen in **Fig. 9**b.

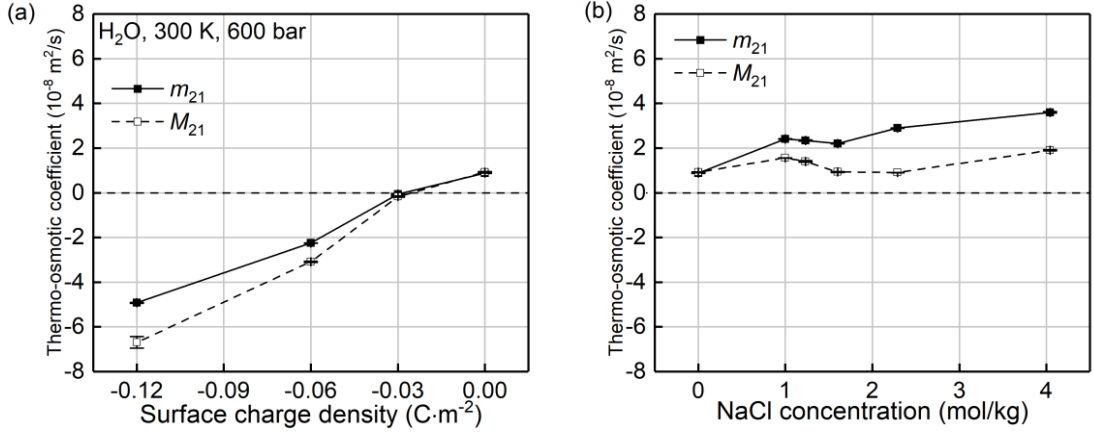

**Fig. 9.** The variations of thermo-osmotic coefficient with (a) surface charge densities, and (b) NaCl concentrations, where $m_{21}$ is computed by Eq. (2) and $M_{21}$ is calculated by Eq. (6).

### 4.5 Electrokinetic properties of nanoconfined liquid

**Surface charge effect.** For system A, using the bulk values for SPC/E water under similar thermodynamic conditions $\varepsilon_b = 70.8\varepsilon_0$[44] and $\eta_b = 0.68 \mathrm{mPa \cdot s}$[51], where $\varepsilon_0$ is the vacuum dielectric permittivity, the computed zeta potential from Eq. (7) is presented as solid symbols in **Fig. 10**a, which shows a monotonic decrease of $\zeta$ potential with decreasing surface charge density from 0 to $-0.12$ C·m$^{-2}$. Since the surface charge density of 0.00, $-0.03$, $-0.06$, and $-0.12$ C·m$^{-2}$, corresponds to the pH values of around 2.0–4.5, 7.5, 9.5, and 11, respectively,[31] the predicted decrease of zeta potential with increasing pH values is consistent with previous experimental and numerical measurements for silica/aqueous electrolyte solution interfaces by Brkljača et al.[54], including the sign inversion at a specific pH value. The calculated values of the zeta potential are also in the same order of magnitude with the experimental observations.[54] A nonzero zeta potential for the uncharged surface is observed in **Fig. 10**a, which has also been reported by the numerical study on the electro-osmotic flow in hydrophobic nanochannels.[28, 29] The calculated $\zeta^*$ from Eq. (8), shown in **Fig. 10**a with hollow symbols, is in the same order of magnitude as $\zeta$. However, it shows a negligible variation with the surface charge density. The discrepancies between $\zeta^*$ and $\zeta$ also come from the linear assumption for the velocity profile in the boundary layers when deriving Eq. (8).

**Concentration effect.** For system B, the zeta potential, measured $\zeta$ and predicted $\zeta^*$, is quantified and shown in **Fig. 10**b. It is found that $\zeta$ linearly and slightly decreases with increasing NaCl concentration, while $\zeta^*$ shows a different variation and larger magnitude. Similarly, the discrepancy is due to the linear assumption for the velocity profiles in the boundary layers when $\zeta^*$ is derived.

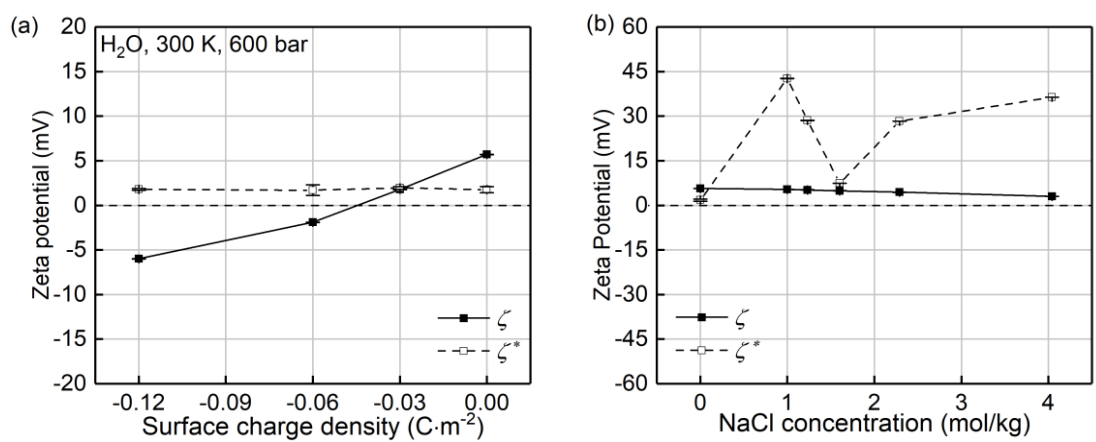

**Fig. 10.** The variations of zeta potential with (a) surface charge densities, and (b) NaCl concentrations, where $\zeta$ is computed by Eq. (7) and $\zeta^*$ is calculated by Eq. (8).

## 5. Discussion

The aim of this section is to decouple the fluxes of solvent and solute based on the results of system B. The movement of nanoconfined ionic solutions due to temperature gradients is not a pure thermo-osmosis problem. It is the result of thermal osmosis combined with thermally induced chemical transport processes (thermal diffusion). Therefore, the thermo-osmotic coefficient quantified in **Fig. 9**b intrinsically includes the Ludwig–Soret effect of ionic species. Based on the thermo-osmotic coefficient measured in this study and thermal diffusion coefficient obtained from a previous one,[15] the fluxes of solvent and solute are de-coupled to assess the extent to which the thermo-osmosis and thermal diffusion co-exist in the nanoconfined fluid mixture when a temperature gradient is imposed. **Fig. 11** shows that thermo-osmosis is the creep fluid flow of interfacial liquid propelled by the pressure gradient induced by the temperature gradient in the boundary layers and this phenomenon will not occur under bulk conditions because thermal gradients do not lead to pressure gradients in the bulk liquid. The thermo-diffusion (i.e., Ludwig–Soret effect) is the separation tendency of different components in interfacial liquid under a temperature gradient, which occurs in both nanoconfined and bulk liquid. In the case of aqueous salt solution, the thermal diffusion is the flux of salt (solute) relative to water molecules (solvent). **Fig. 11**b shows that the thermally induced solution flux in a nanochannel is the combination of thermo-osmotic flow and thermo-diffusive flow. Therefore, the thermo-osmotic coefficient quantified in our study includes the Ludwig–Soret effect of NaCl ions.

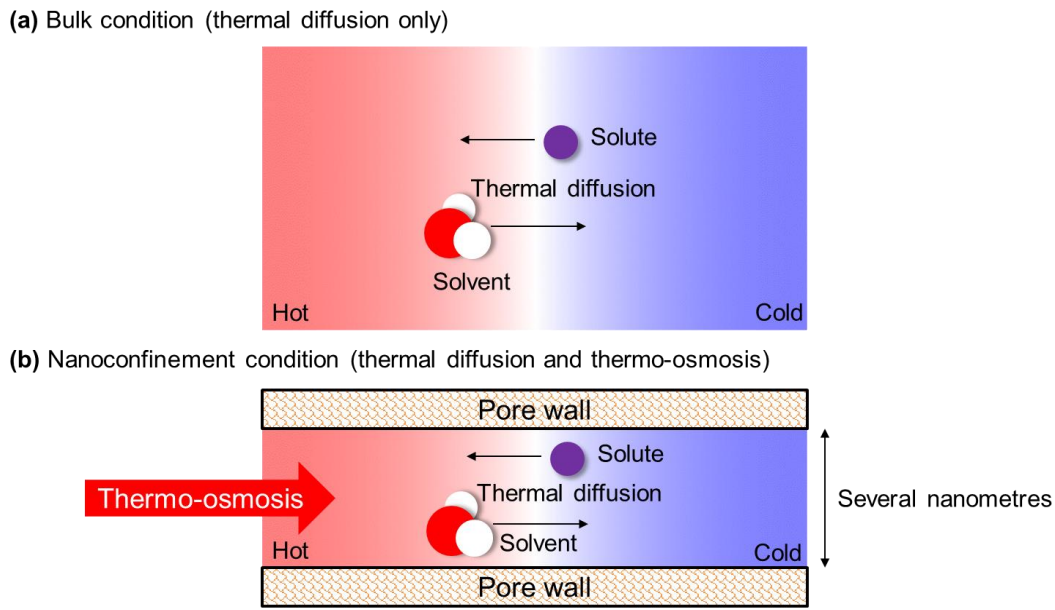

**Fig. 11.** Thermally induced mass transport phenomena including thermo-osmosis and thermal diffusion, occurring under (a) bulk and (b) nanoconfined conditions.

In the nonequilibrium thermodynamics and in the linear regimes, the flux of nanoconfined solution, $J_{\text{Solution}}$, (volumetric flow rate, unit: m$^3$/s) under a temperature gradient and without the pressure and chemical concentration gradients can be described by:

$$J_{\text{Solution}} = \bar{V}_{\text{Solvent}} J_{\text{Solvent}} + \bar{V}_{\text{Solute}} J_{\text{Solute}} = \left(-m_{21} \frac{\nabla T}{T}\right) A \tag{11}$$

where, $\bar{V}_{\text{Solvent}}$ and $\bar{V}_{\text{Solute}}$ are the molar volume of solvent and solute, $J_{\text{Solvent}}$ and $J_{\text{Solute}}$ the fluxes of solvent and solute, $m_{21}$ the thermo-osmotic coefficient of the solution as obtained in this study, $A$ the cross section for the solution flux.

The excess solute flux (unit: mol/s, the solute flux relative to the solvent flux, computed as the difference between the observed solute flux $J_{\text{Solute}}$ and solute flux predicted according to the total solute concentration, $c_{\text{Solute}}$, and solution flux, $J_{\text{Solution}}$) is induced by the thermal diffusion:

$$J_{\text{Solute}} - c_{\text{Solute}} J_{\text{Solution}} = -\rho x_{\text{Solute}}(1 - x_{\text{Solute}})(D_T T)\frac{\nabla T}{T} \tag{12}$$

where, $\rho$ is the molar density of the mixture, $x_{\text{Solute}}$ the molar fraction of the solute, $D_T T$ the thermal diffusivity of the solute.

Eq. (11) and Eq. (12) show that the solute flux and solvent flux can be determined if all other data are known. Taking our case with the NaCl concentration of 4.04 mol/kg, the surface charge density of $0.00 \text{C} \cdot \text{m}^{-2}$, and a unit temperature gradient as an example, the corresponding data from this study and the previous one on the thermal diffusion[15] in the same nanofluidic system is presented in **Table 1**. The obtained $J_{\text{Solvent}}$ and $J_{\text{Solute}}$ indicate that the thermally induced migration of water molecules and dissolved NaCl ions are in opposite directions due to the thermophilic thermal diffusion of NaCl ions, i.e., $D_T T < 0$.

Table 1. The input data to decouple the solvent flux and solute flux and the obtained results.

| Variable | Value | Unit | Source |
|---|---|---|---|
| $\bar{V}_{Solvent}$ | 1.80797 | $10^{-5} m^3$/mol | This study using the Voronoi volume averaged over all solvent particles |
| $\bar{V}_{Solute}$ | 1.01293 | $10^{-5} m^3$/mol | This study using the Voronoi volume averaged over all solute particles |
| $m_{21}$ | 3.5988 | $10^{-8} m^2/s$ | This study |
| $A$ | 10.4674 | $10^{-18} m^2$ | This study |
| $c_{Solute}$ | 7635.971969 | $mol/m^3$ | This study |
| $\rho$ | 47782.67308 | $mol/m^3$ | This study |
| $x_{Solute}$ | 0.1598 | 1 | This study |
| $D_T T$ | -0.01646 | $10^{-9} m^2/s$ | Ref. [15] |
| $-\dfrac{\nabla T}{T}$ | 1 | $m^{-1}$ | This study |
| $J_{Solvent}$ | 5.91627 | $10^{-8} mol/s$ | Calculated |
| $J_{Solute}$ | -10.5599 | $10^{-8} mol/s$ | Calculated |

## 6. Conclusions

MD simulations were used to investigate the thermo-osmotic response of pure water confined in charged silica slit nanochannels with surface charge density varying from $0.00\ C \cdot m^{-2}$ to $-0.12\ C \cdot m^{-2}$. The NEMD results showed that decreasing surface charges reduces progressively the self-diffusivity and thermo-osmotic mobility of the nanoconfined liquid. The thermo-osmotic flow was found to change direction for surface charge densities less than $-0.03 C \cdot m^{-2}$. It was shown that the structural modifications caused by the nanoconfinement, and the surface charge were significantly correlated with the quantified thermo-osmotic properties of the confined liquid. In addition, by further dissolving NaCl ions to the interfacial water confined by uncharged silica nanochannels, the ion concentration effect was found to facilitate the interfacial thermo-osmotic flow. The fluxes of solvent and solute in the thermo-osmotic flow were decoupled by considering the Ludwig–Soret effect of NaCl ions. It was shown that these two fluxes were in opposite directions due to the thermophilic thermal diffusion of NaCl ions.

The presented study was carried out on realistic surfaces. It is the first one to reveal that the thermo-osmotic response of nanoconfined liquid can be manipulated by changing the surface charge density and ionic strength. The study provides molecular-level quantifications and explanations for the existing experimental observations and macroscopic analysis. The simulation methods adopted in this work and the insights provided by the results, form a methodology for further research on coupled transport phenomena in nanostructures.

**Declaration of Competing Interest**

The authors declare that they have no known competing financial interests or personal relationships that could have appeared to influence the work reported in this paper.

**Acknowledgment**

Chen acknowledges the President Doctoral Scholarship Award (PDS Award 2019) by The University of Manchester, UK. Jivkov acknowledges gratefully the financial support from the Engineering and Physical Sciences Research Council (EPSRC), UK, via Grant EP/N026136/1. The authors acknowledge the assistance provided by the Research IT team for the use of Computational Shared Facility at The University of Manchester.